\documentclass{article}
\usepackage{amssymb}

\usepackage{graphicx}
\usepackage{amsmath}
\usepackage{epstopdf}
\usepackage[utf8]{inputenc}
\usepackage[T1]{fontenc}
\usepackage{mathrsfs}

\newtheorem{theorem}{Theorem}

\newtheorem{proposition}[theorem]{Proposition}

\begin{document}

\title{Belavkin Filtering with Squeezed Light Sources}
\author{Anita D\k{a}browska$^{1}$, John Gough$^{2}$ \\
%EndAName
1. Department of Theoretical Foundations of Biomedical Science\\
and Medical Informatics, ul. Jagiello\'{n}ska 15, 85-067 Bydgoszcz,\\
Nicolaus Copernicus University, Poland\\
Email: \texttt{adabro@cm.umk.pl} \\
2. Department of Mathematics and Physics,\\
Aberystwyth University, SY23 3BZ, UK.\\
Email: \texttt{jug@aber.ac.uk.}}
\date{}
\maketitle

\begin{abstract}
We derive the filtering equation for Markovian systems undergoing homodyne
measurement in the situation where the output processes being monitored are
squeezed. The filtering theory applies to case where the system is driven by
Fock noise (that, quantum input processes in a coherent state) and where the
output is mixed with a squeezed signal. It also applies to the case of a
system driven by squeezed noise, but here there is a physical restriction to
emission/absorption coupling only. For the special case of a cavity mode
where the dynamics is linear, we are able to derive explicitly the filtered
estimate $\pi_t (a)$ for the mode annihilator $a$ based on the homodyne
quadrature observations up to time $t$.
\end{abstract}

\begin{center}
In memory of Slava Belavkin.
\end{center}

\section{Introduction}

\label{sec:intro}

The theory of quantum filtering was developed by V.P. Belavkin in the 1980's 
\cite{Bel80}-\cite{Bel89} as the extension of classical filtering theory 
\cite{Strat59}-\cite{DavisMarcus}. It has been subsequently developed as a
technique for quantum measurement and control \cite{Bar90}-\cite
{nonclassical}.

\begin{figure}[tbph]
\centering
\includegraphics[width=1.00\textwidth]{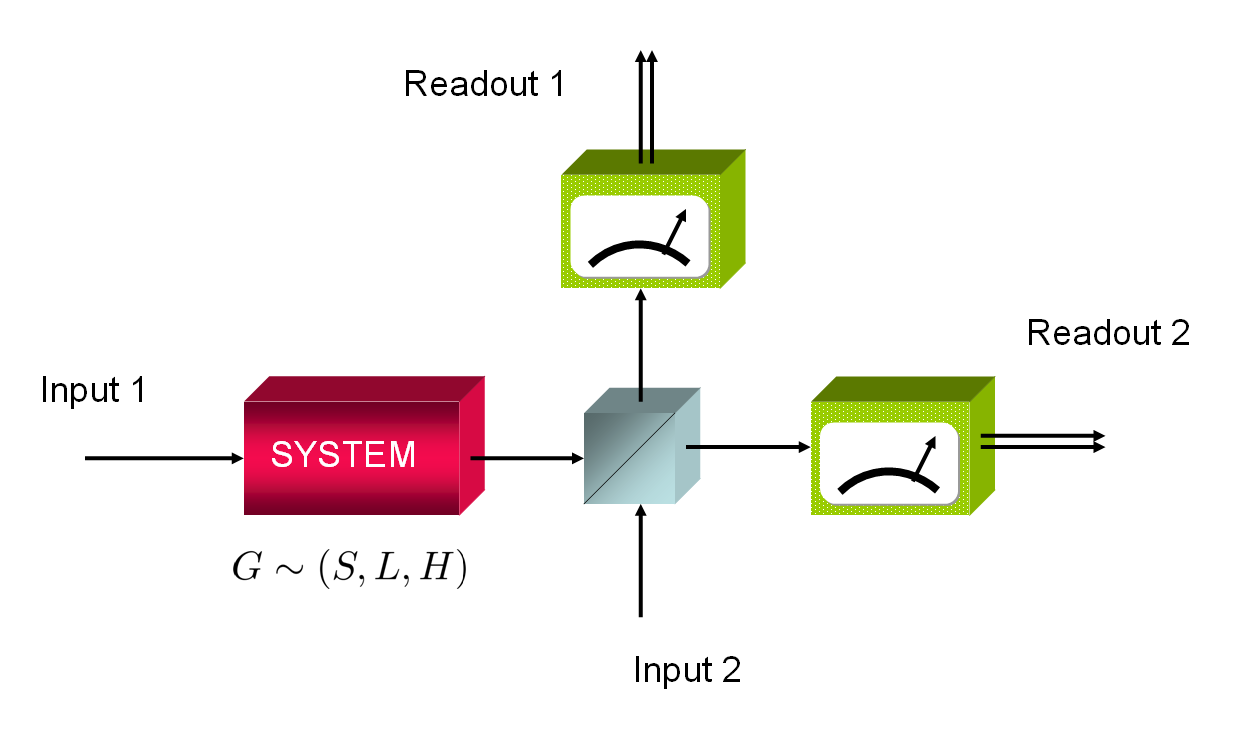}
\caption{Experimental setup: an open quantum system is driven by a coherent
state field (input 1) and the output enters a 50-50 beamsplitter along with
a squeezed coherent state field (input 2). The single-line arrows stand for
quantum field inputs, while the double-line arrows imply classical readout.}
\label{fig:squeezed_input_filter}
\end{figure}

Our aim here is to derive the filter for a system whose output is either
squeezed light, or is mixed with squeezed light. For the special case of a
linear dynamics, we are able to give the filter explicitly. The advantage of
using squeezing as a control resource was proposed in \cite{YK1} and \cite
{YK2}. In the experimental setup in figure \ref{fig:squeezed_input_filter}
below we have a input-system-output device with dynamical coupling operators 
$\left( S,L,H\right) $, see later, whose output is sent through a beam
splitter and mixed with a second input. The input into the system will be
modelled as a coherent state, however we take the second input to be in a
squeezed coherent state. The output processes from the beam splitter are
then subject to a homodyne measurement. The scheme depicted in figure \ref
{fig:squeezed_input_filter} has been proposed in \cite{hofer11} as a
component of an entanglement with large squeezing to suppress noise.

The input processes are described by independent quantum Wiener annihilation
processes $A(\cdot )$ and $B\left( \cdot \right) $ and we have 
\begin{equation*}
\left[ A\left( t\right) ,A^{\ast }\left( s\right) \right] =\left[ B\left(
t\right) ,B^{\ast }\left( s\right) \right] =t\wedge s\,,
\end{equation*}
where $t\wedge s=\mathrm{min}(t,s)$. The inputs are in Gaussian states with
means 
\begin{equation}
\mathbb{E}\left[ dA\left( t\right) \right] =\alpha \left( t\right) \,dt\quad
,\left[ dB(t)\right] =\beta \left( t\right) .  \label{eq:means}
\end{equation}
We shall take the two inputs to be independent, with input one being a
coherent state (and so having the same covariances as the vacuum) while
input two is squeezed. The quantum It\={o} table \cite{HP} will then have
the form 
\begin{equation}
\begin{tabular}{l|ll}
$\times $ & $dB$ & $dB^{\ast }$ \\ \hline
$dB$ & 0 & $dt$ \\ 
$dB^{\ast }$ & 0 & 0
\end{tabular}
,\quad 
\begin{tabular}{l|ll}
$\times $ & $dA$ & $dA^{\ast }$ \\ \hline
$dA$ & $mdt$ & $\left( n+1\right) dt$ \\ 
$dA^{\ast }$ & $ndt$ & $m^{\ast }dt$%
\end{tabular}
,  \label{table}
\end{equation}
where $n>1$ and $\left| m\right| ^{2}\leq n\left( n+1\right) $. Note that $%
dAdB^{\ast }\equiv 0$, etc.

The output $B^{\text{out}}$ from the system is passed through the beam
splitter along with the squeezed input $A^{\text{out}}\equiv A$ leading to
the overall output 
\begin{equation*}
\left[ 
\begin{array}{c}
C_{1} \\ 
C_{2}
\end{array}
\right] =\frac{1}{\sqrt{2}}\left[ 
\begin{array}{cc}
1 & 1 \\ 
-i & i
\end{array}
\right] \left[ 
\begin{array}{c}
B^{\text{out}} \\ 
A^{\text{out}}
\end{array}
\right]
\end{equation*}
with $C_{1}=\frac{1}{\sqrt{2}}\left( B^{\text{out}}+A^{\text{out}}\right) $
and $C_{2}=\frac{1}{\sqrt{2}i}\left( B^{\text{out}}-A^{\text{out}}\right) $
then being independent annihilation processes: $\left[ C_{j}\left( t\right)
,C_{k}^{\ast }\left( s\right) \right] =\delta _{jk}\,\left( t\wedge s\right) 
$.

Finally we measure the quadratures 
\begin{eqnarray*}
Y_{1} &=&C_{1}+C_{1}^{\ast }\equiv \frac{1}{\sqrt{2}}\left( Q_{B}^{\text{out}%
}+Q_{A}^{\text{out}}\right) \\
Y_{2} &=&C_{2}+C_{2}^{\ast }\equiv \frac{1}{\sqrt{2}}\left( P_{B}^{\text{out}%
}-P_{A}^{\text{out}}\right)
\end{eqnarray*}
at the detectors. Here $Q_{B}^{\text{out}}=B^{\text{out}}+B^{\text{out}\ast
} $ and $P_{B}^{\text{out}}=\frac{1}{i}(B^{\text{out}}-B^{\text{out}\ast })$%
, etc. We will see that quadratures then have increments of the form 
\begin{eqnarray}
dY_{1}\left( t\right) &=&\frac{1}{\sqrt{2}}\left\{ j_{t}\left( S\right)
dB\left( t\right) +j_{t}\left( L\right) dt+dA\left( t\right) +\text{H.c.}%
\right\} ,  \notag \\
dY_{2}\left( t\right) &=&\frac{1}{\sqrt{2}i}\left\{ j_{t}\left( S\right)
dB\left( t\right) +j_{t}\left( L\right) dt-dA\left( t\right) -\text{H.c.}%
\right\} ,  \label{eq:Y-Example}
\end{eqnarray}
where $j_{t}\left( \cdot \right) $ transfers to the interaction picture, and 
$S$ and $L$ are prescribed operators determining how the system couples to
the input field.

The outputs $Y_{1}$ and $Y_{2}$ have non-trivial correlation expressed
through the following table 
\begin{equation*}
\begin{tabular}{l|ll}
$\times $ & $dY_{1}$ & $dY_{2}$ \\ \hline
$dY_{1}$ & $\left( 1+n+m^{\prime }\right) dt$ & $m^{\prime \prime }dt$ \\ 
$dY_{2}$ & $m^{\prime \prime }dt$ & $\left( 1+n-m^{\prime }\right) dt$%
\end{tabular}
\end{equation*}
where $m=m^{\prime }+im^{\prime \prime }$ is the decomposition of the
squeezing parameter into real and imaginary parts. In particular, the matrix 
\begin{equation}
K=\left[ 
\begin{array}{cc}
1+n+m^{\prime } & m^{\prime \prime } \\ 
m^{\prime \prime } & 1+n-m^{\prime }
\end{array}
\right]  \label{eq: K}
\end{equation}
has determinant $\Delta =\left( 1+n\right) ^{2}-\left| m\right| ^{2}>1+n$
and therefore is invertible.

Our interest will be in modelling the dynamical evolution of the system
conditioned on the continuous observations of the output quadratures $Y_{1}$
and $Y_{2}$.

\section{Quantum Filtering for General Gaussian States}

In this section we derive the filter for a generic quantum Markov model with
inputs in Gaussian states. Generalizing the situation in the introduction,
we allow for $m_{\text{Fock}}$ vacuum inputs $B_{1},\cdots ,B_{m_{\text{Fock}%
}}$ and $m_{\text{sq}}$ squeezed inputs $A_{1},\cdots ,A_{m_{\text{sq}}}$.
We may think of the input driving a single component (with Hilbert space $%
\mathfrak{h}$) which in fact is a network of several quantum inputs. The
inputs are fields on a Fock space $\mathfrak{F}$. The outputs are then fed
into a measurement apparatus which performs a continuous measurement on a
collection $Y_{1},\cdots ,Y_{n_{\text{obs}}}$ of commuting self-adjoint
processes. We shall assume that the measured processes are linear
combinations of the output quadratures, and that they are linearly
independent (so $n_{\text{obs}}\leq m_{\text{Fock}}+m_{\text{sq}}$). The
set-up is depicted in Figure 2.

\begin{figure}[tbph]
\centering
\includegraphics[width=1.00\textwidth]{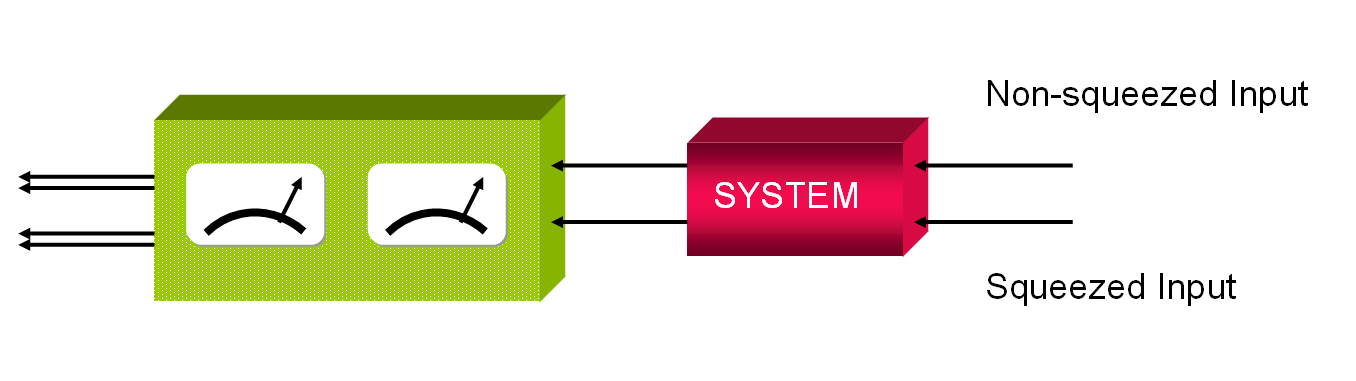}
\caption{Equivalent block-diagram set-up of the system in figure 1.}
\label{fig:squeezed_model}
\end{figure}

On the joint space $\mathfrak{h}\otimes \mathfrak{F}$, we consider the
quantum stochastic process $V(\cdot )$ satisfying the quantum stochastic
differential equation (QSDE) \cite{HP} 
\begin{eqnarray}
dV(t) &=&\{\left( S_{jk}-\delta _{jk}\right) \otimes d\Lambda _{jk}(t) 
\notag \\
&&+L_{j}\otimes dB_{j}^{\ast }\left( t\right) +R_{j}\otimes dA_{j}^{\ast } 
\notag \\
&&-L_{j}^{\ast }S_{jk}\otimes dB_{k}\left( t\right) -R_{j}^{\ast }\otimes
dA_{j}  \notag \\
&&-(K_{A}+K_{B}+iH)\otimes dt\}\,V(t)  \label{eq:unitary_QSDE}
\end{eqnarray}
with $V\left( 0\right) =1$.

Here $S=\left[ S_{jk}\right] $ is a unitary $m_{\text{Fock}}\times m_{\text{%
Fock}}$ matrix whose entries are bounded operators on $\mathfrak{h}$, $L_{j}$
$\left( j=1,\cdots ,m_{\text{Fock}}\right) $ and $R_{j}$ $\left( j=1,\cdots
,m_{\text{sq}}\right) $ are bounded operators and $H$ self-adjoint. This
specific form of QSDE may be termed the \emph{Hudson-Parthasarathy equation}
as the algebraic conditions on the coefficients are necessary and sufficient
to ensure unitarity (though the restriction of boundedness can be lifted).
The process is also adapted in the sense of Hudson and Parthasarathy \cite
{HP}. The operators $K_A$ and $K_B$ will be given below as (\ref{eq:Ks})
and a representation specific.

The processes $\Lambda _{jk}\left( t\right) $ are the scattering (or gauge)
processes. Formally we may introduce $b_{j}(t)$ as the derivative of $%
B_{j}(t)$ in which case 
\begin{equation*}
\Lambda _{jk}\left( t\right) \equiv \int_{0}^{t}b_{j}^{\ast }(\tau
)b_{k}(\tau )\,d\tau .
\end{equation*}
We adopt the summation convention that repeated Latin indices are summed
from 1 to $m_{\text{Fock}}$ for the non-squeezed terms (e.g. $B_{j},\Lambda
_{jk},L_{j},S_{jk}$), and from 1 to $m_{\text{sq}}$ for the squeezed terms
(e.g. $A_{j},R_{j}$). The ranges should be clear from the context. \textbf{%
Note that the squeezed terms have no scattering terms!} The squeezed terms
rely on the non-Fock (i.e., Araki-Woods) representation and the scattering
process is not well-defined in this case.

\subsection{Gaussian Input States}

We shall aim for the most general Gaussian state for the input field. Let us
introduce the Weyl displacement operator 
\begin{eqnarray*}
W(g,f) &=&\exp [\int_{0}^{\infty }g_{k}(t)dB_{k}(t)^{\ast
}+f_{j}(t)dA_{j}(t)^{\ast } \\
&&-g_{k}\left( t\right) ^{\ast }dB_{k}\left( t\right) -f_{j}\left( t\right)
^{\ast }dA_{j}\left( t\right) ]
\end{eqnarray*}
with square-integrable test functions $g_{k},f_{k}$. We require a Gaussian
state where the $A$ fields are squeezed and the $B$ fields are unsqueezed
(Fock): here we take 
\begin{eqnarray*}
\mathbb{E}\left[ W\left( g,f\right) \right] &=&\exp \{\int_{0}^{\infty }%
\left[ g_{k}(t)\beta _{k}(t)^{\ast }-g_{k}\left( t\right) ^{\ast }\beta
_{k}\left( t\right) \right] dt-\frac{1}{2}\int g_{k}^{\ast }(t)g_{k}(t)dt\}
\\
&&\times \exp \{\int_{0}^{\infty }\left[ f_{k}(t)\alpha _{k}(t)^{\ast
}-f_{k}\left( t\right) ^{\ast }\alpha _{k}\left( t\right) \right] dt \\
&&-\frac{1}{2}\int_{0}^{\infty } 
\begin{array}{cc}
\lbrack \vec{f}^{\prime }(t)^{\top }, & -\vec{f}^{\prime \prime }(t)^{\top }]
\\ 
\, & \,
\end{array}
C\left[ 
\begin{array}{c}
\vec{f}^{\prime }(t) \\ 
\vec{f}^{\prime \prime }(t)
\end{array}
\right] dt\}.
\end{eqnarray*}
Here the means are $\alpha _{j}$ are $\beta _{j}$ giving the
multi-dimensional version of (\ref{eq:means}). The $A$ quadratures and $B$
quadratures are always independent for this state. We also have $\vec{f}%
^{\prime }$ and $\vec{f}^{\prime \prime }$ as the vectors whose entries are
the real and imaginary components of the fields $f_{j}$. Introducing the
quadratures 
\begin{eqnarray*}
Q_{k} &=&A_{k}+A_{k}^{\ast }, \\
P_{k} &=&\frac{1}{i}\left( A_{k}-A_{k}^{\ast }\right) ,
\end{eqnarray*}
and set $\left[ X_{j}\left( t\right) \right]_1^{2 m_{\text{sq}} } =\left(
Q_{1}\left( t\right) ,P_{1}\left( t\right) ,\cdots ,Q_{m_{\text{sq}}}\left(
t\right) ,P_{m_{\text{sq}}}\left( t\right) \right) $. Then 
\begin{equation*}
\left[ X_{j}\left( t\right) ,X_{k}\left( s\right) \right] =2iJ_{jk}\,\min
\left\{ t,s\right\}
\end{equation*}
where $J=\oplus _{j=1}^{m_{\text{sq}}}\left[ 
\begin{array}{cc}
0 & 1 \\ 
-1 & 0
\end{array}
\right] $. The covariance matrix $C$ is then 
\begin{equation*}
C_{jk}=\text{Re }\mathbb{E}\left[ \left( X_{j}-\bar{X}_{j}\right) \left(
X_{k}-\bar{X}_{k}\right) \right]
\end{equation*}
which is a $\left( 2m_{\mathrm{sq}} \right) $-square symmetric matrix. The
Heisenberg uncertainty principle implies that $C$ cannot be chosen
arbitrarily, but must satisfy 
\begin{equation*}
C+i\,J\geq 0.
\end{equation*}
We may write 
\begin{multline*}
\mathbb{E}\left[ W\left( g,f\right) \right] =\exp \{2\int_{0}^{\infty }\left[
g_{k}^{\prime \prime }(t)\beta _{k}^{\prime }(t)^{\ast }-g_{k}^{\prime
}\left( t\right) ^{\ast }\beta _{k}^{\prime \prime }\left( t\right) \right]
dt \\
-\frac{1}{2}\int_{0}^{\infty }[g_{j}^{\prime }\left( t\right) g_{j}^{\prime
}\left( t\right) +g_{j}^{\prime \prime }\left( t\right) g_{j}^{\prime \prime
}\left( t\right) ]dt \\
+2\int_{0}^{\infty }\left[ f_{k}^{\prime \prime }(t)\alpha _{k}^{\prime
}(t)^{\ast }-f_{k}^{\prime }\left( t\right) ^{\ast }\alpha _{k}^{\prime
\prime }\left( t\right) \right] dt \\
-\frac{1}{2}\int_{0}^{\infty }[f_{j}^{\prime }\left( t\right) f_{k}^{\prime
}\left( t\right) C_{jk}^{PP}+f_{j}^{\prime \prime }\left( t\right)
f_{k}^{\prime \prime }\left( t\right) C_{jk}^{QQ}+2f_{j}^{\prime }\left(
t\right) f_{k}^{\prime \prime }\left( t\right) C_{jk}^{QP}]\}.
\end{multline*}

\subsubsection{The Quantum It\={o} Table}

The quantum It\={o} table for the squeezed fields takes the form 
\begin{eqnarray*}
dA_{j}dA_{k}^{\ast } &=&\left( \delta _{jk}+N_{kj}\right) \,dt \\
dA_{j}^{\ast }dA_{k} &=&N_{jk}\,dt \\
dA_{j}dA_{k} &=&M_{jk}\,dt \\
dA_{j}^{\ast }dA_{k}^{\ast } &=&M_{kj}^{\ast }\,dt
\end{eqnarray*}
where $N=N^{\ast }$ and $M=M^{\top }$ are complex $m_{\text{sq}}\times m_{%
\text{sq}}$ matrices. The matrices $N,M$ are fully determined from the
covariance $C$ and vice versa. From the relation $C_{jk}^{QQ}dt=dQ_{j}dQ_{k}$
we see that 
\begin{equation*}
dQ_{j}dQ_{k}=C_{jk}^{QQ}\,dt
\end{equation*}
with 
\begin{equation*}
C^{QQ}=I_{n_{\text{obs}}}+N+N^{\top }+M+M^{\ast }.
\end{equation*}
This generalizes (\ref{eq: K}) to the multidimensional case.

For the unsqueezed fields, we may additionally include the scattering
process to get 
\begin{eqnarray*}
dB_{j}dB_{k}^{\ast } &=&\delta _{jk}\,dt \\
dB_{j}d\Lambda _{kl} &=&\delta _{jk}\,dB_{l} \\
d\Lambda _{jk}dB_{l}^{\ast } &=&\delta _{kl}\,dB_{j}^{\ast } \\
d\Lambda _{jk}d\Lambda _{lr} &=&\delta _{kl}\,d\Lambda _{jr}.
\end{eqnarray*}
All other product of the increments $dt,dA_{j},dA_{j}^{\ast
},dB_{k},dB_{k}^{\ast },d\Lambda _{jl}$ vanish.

\subsection{The Heisenberg-Langevin Equations}

The quantum stochastic process defined by (\ref{eq:unitary_QSDE}) defines a
unitary if and only if $S=\left[ S_{jk}\right] $ is unitary, $H$ is
self-adjoint, and 
\begin{eqnarray}
K_{B} &=&\frac{1}{2}L_{k}^{\ast }L_{k}, \nonumber \\
K_{A} &=&\frac{1}{2}R_{j}^{\ast }\left( \delta _{jk}+N_{kj}\right) R_{k}+
\frac{1}{2}R_{j}N_{jk}R_{k}^{\ast } \nonumber \\
&&-\frac{1}{2}R_{j}^{\ast }M_{jk}R_{k}^{\ast }-\frac{1}{2}R_{j}M_{jk}^{\ast
}R_{k}.
\label{eq:Ks}
\end{eqnarray}

For a given system operator $X$ we set 
\begin{equation}
j_{t}\left( X\right) \triangleq V^{\ast }\left( t\right) \left[ X\otimes I%
\right] V\left( t\right) .
\end{equation}
Then from the quantum It\={o} calculus we get 
\begin{eqnarray}
dj_{t}\left( X\right) &=&j_{t}\left( \mathcal{L}_{jk}X\right) \otimes
d\Lambda _{jk}(t)  \notag \\
&&+j_{t}\left( \mathcal{L}_{j0}X\right) \otimes dB_{j}^{\ast }\left(
t\right) +j_{t}\left( \mathcal{R}_{j0}X\right) \otimes dA_{j}^{\ast }\left(
t\right)  \notag \\
&&+j_{t}\left( \mathcal{L}_{0k}X\right) \otimes dB_{k}\left( t\right)
+j_{t}\left( \mathcal{R}_{0k}X\right) \otimes dA_{k}\left( t\right)  \notag
\\
&&+j_{t}\left( \mathcal{L}_{00}X+\mathcal{R}_{00}X-i\left[ X,H\right]
\right) \otimes dt  \label{eq:dynamical}
\end{eqnarray}
where the Evans-Hudson maps $\mathcal{L}_{\mu \nu }$ are explicitly given by 
\begin{eqnarray*}
\mathcal{L}_{jk}X &=&S_{lj}^{\ast }XS_{lk}-\delta _{jk}X, \\
\mathcal{L}_{j0}X &=&S_{lj}^{\ast }[X,L_{l}], \\
\mathcal{L}_{0k}X &=&[L_{l}^{\ast },X]S_{lk} \\
\mathcal{L}_{00}X &=&\mathcal{L}_{L}X
\end{eqnarray*}
and 
\begin{eqnarray*}
\mathcal{R}_{j0}X &=&[X,R_{j}], \\
\mathcal{R}_{0k}X &=&[R_{k}^{\ast },X], \\
\mathcal{R}_{00}X &=&\mathcal{L}_{R}X+\frac{1}{2}N_{jk}\left\{ R_{k}^{\ast
}[X,R_{j}]+[R_{k}^{\ast },X]R_{j}\right\} \\
&&+\frac{1}{2}N_{kj}\left\{ R_{k}[X,R_{j}^{\ast }]+[R_{k},X]R_{j}^{\ast
}\right\} \\
&&-\frac{1}{2}M_{jk}^{\ast}\left\{ R_{k}[X,R_{j}]+[R_{k},X]R_{j}\right\} \\
&&-\frac{1}{2}M_{jk}\left\{ R_{k}^{\ast }[X,R_{j}^{\ast
}]+[R_{k}^{\ast },X]R_{j}^{\ast }\right\}
\end{eqnarray*}
where we have the standard pure Lindblad superoperator 
\begin{equation}
\mathcal{L}_{L}X\triangleq \frac{1}{2}L_{l}^{\ast }[X,L_{l}]+\frac{1}{2}[%
L_{l}^{\ast },X]L_{l}.
\end{equation}

\subsection{Output Processes}

We introduce the processes 
\begin{eqnarray}
A_{j}^{\text{out}}\left( t\right) &\triangleq &V^{\ast }\left( t\right) %
\left[ I\otimes A_{j}\left( t\right) \right] V\left( t\right)  \notag \\
B_{j}^{\text{out}}\left( t\right) &\triangleq &V^{\ast }\left( t\right) %
\left[ I\otimes B_{j}\left( t\right) \right] V\left( t\right) .
\end{eqnarray}
We note that we equivalently have $A_{k}^{\text{out}}\left( t\right) \equiv
V^{\ast }\left( T\right) \left[ 1\otimes A_{k}\left( t\right) \right]
V\left( T\right) $, for $t\leq T$, and similarly for the $B$ fields. Using
the quantum It\={o} rules, we see that 
\begin{eqnarray}
dA_{k}^{\text{out}} &=&dA_{k}+j_{t}(R_{k})dt,  \notag \\
dB_{k}^{\text{out}} &=&j_{t}(S_{kl})dB_{l}(t)+j_{t}(L_{k})dt.
\label{eq:output}
\end{eqnarray}

The readout will consist of the observations of processes $\left( \alpha
=1,\cdots ,n_{\text{obs}}\right) $ 
\begin{eqnarray*}
Y_{\alpha }\left( t\right) &=&T_{\alpha j}B_{j}^{\text{out}}\left( t\right)
+U_{\alpha k}A_{k}^{\text{out}}\left( t\right) +\text{H.c.} \\
&\equiv &V^{\ast }\left( t\right) \left[ I\otimes \left\{ T_{\alpha
j}B_{j}\left( t\right) +U_{\alpha k}A_{k}\left( t\right) +\text{H.c.}%
\right\} \right] V\left( t\right) ,
\end{eqnarray*}
where the $T_{\alpha k}$ and $U_{\alpha j}$ are complex constants. We
require that the $Y_{\alpha }\left( t\right) $ commute for all $\alpha $ and 
$t\geq 0$. This requires the following identity $Z_{\alpha \beta }=Z_{\beta
\alpha }$ where we introduce 
\begin{equation}
Z_{\alpha \beta }\triangleq T_{\alpha k}T_{\beta k}^{\ast }+U_{\alpha
j}U_{\beta k}^{\ast }.  \label{eq:Z}
\end{equation}

We will have 
\begin{equation}
dY_{\alpha }= j_{t}\left( T_{\alpha j}S_{jk}\right) \,dB_{k}+ U_{\alpha k}
dA_{k}+j_{t}(T_{\alpha j}L_{j}+U_{\alpha j} R_j)\,dt+\text{H.c.}
\label{eq:dY}
\end{equation}
We now consider the problem of continuously measuring processes $%
Y_{1},\cdots ,Y_{n_{\text{obs}}}$. These generate the measurement algebra up
to time $t$: 
\begin{equation}
\mathfrak{Y}_{t]}=\mathrm{vN}\left\{ Y_{\alpha }\left( s\right) :\alpha
=1,\cdots ,n_{\text{obs}},\,0\leq s\leq t\right\} .
\end{equation}
The family $\left\{ \mathfrak{Y}_{t]}:t\geq 0\right\} $ then forms an
increasing family (\textit{filtration)} of commutative von Neumann algebras.

The correlation matrix $K=[K_{\alpha \beta }]$ is defined by 
\begin{equation}
dY_{\alpha }\,dY_{\beta }=K_{\alpha \beta }\,dt  \label{eq:correlation}
\end{equation}
From the increment (\ref{eq:dY}) we see that 
\begin{equation}
K_{\alpha \beta }\equiv Z_{\alpha \beta }+U_{\alpha k}N_{jk}U_{\beta
j}^{\ast }+U_{\alpha k}^{\ast }N_{jk}U_{\beta j}+U_{\alpha k}M_{jk}U_{\beta
j}+U_{\alpha k}^{\ast }M_{jk}^{\ast }U_{\beta j}^{\ast }.
\label{eq:K_from_ZNMU}
\end{equation}

\subsection{The Filter}

The filtered estimate for $j_{t}\left( X\right) $ given the measurement
readout is then 
\begin{equation*}
\pi _{t}\left( X\right) :=\mathbb{E}\left[ j_{t}\left( X\right) \mid %
\mathfrak{Y}_{t]}\right] ,
\end{equation*}
where $\mathbb{E}\left[ \cdot \mid \mathfrak{Y}_{t]}\right] $ is the
conditional expectation onto the measurement readout algebra up to time $t$
for the given state $\mathbb{E}$ which is the product state for the system
with the Gaussian states for the fields. The conditional expectation has the
least squares property that 
\begin{equation*}
\mathbb{E}\left[ \left\{ \hat{X}_{t}-j_{t}\left( X\right) \right\} ^{2}%
\right]
\end{equation*}
is a minimum over all $\hat{X}_{t}\in \mathfrak{Y}_{t]}$ for the choice $%
\hat{X}_{t}=\pi _{t}\left( X\right) $. This implies the condition that 
\begin{equation*}
\mathbb{E}\left[ \left\{ \pi _{t}\left( X\right) -j_{t}\left( X\right)
\right\} C\left( t\right) \right] =0
\end{equation*}
for every $C\left( t\right) \in \mathfrak{Y}_{t]}$.

We now state the main result which we derive in the next subsection. (We
shall adopt the convention that repeated Greek indices implies a sum over
the range 1 to $n_{\text{obs}}$.)

\begin{theorem}
\label{thm:filter} The filter satisfies the Belavkin-Kushner-Stratonovich
equation 
\begin{equation}
d\pi _{t}(X)=\pi _{t}(\mathcal{L}X)\,dt+\mathcal{H}_{t}^{\alpha }\left(
X\right) dI_{\alpha }\left( t\right) ,
\end{equation}
where (suppressing the time index $t$) $\mathcal{L}X=\tilde{\mathcal{L}}_{00}X+\mathcal{\tilde{R}}_{00}X-i%
\left[ X,H\right] $ with 
\begin{eqnarray*}
\tilde{\mathcal{L}}_{00}X &=&\mathcal{L}_{00}X+\beta _{j}^{\ast }(t)\mathcal{%
L}_{j0}X+\mathcal{L}_{0k}X\beta _{k}(t)+\beta _{j}^{\ast }\mathcal{L}%
_{jk}X\beta _{k}(t), \\
\mathcal{\tilde{R}}_{00}X &=&\mathcal{R}_{00}X+\alpha _{j}^{\ast }(t)%
\mathcal{R}_{j0}X+\mathcal{R}_{0k}X\alpha _{k}(t),
\end{eqnarray*}
and 
\begin{eqnarray}
\mathcal{H}_{t}^{\alpha }\left( X\right) &=&\pi _{t}(X\tilde{L}^{\alpha }+%
\tilde{L}^{\alpha \ast }X)-\pi _{t}(X) \pi_t(\tilde{L}^{\alpha }+\tilde{L}%
^{\alpha \ast })  \notag \\
&&+\pi _{t}([X,\tilde{R}^{\alpha }])+\pi _{t}([\tilde{R}^{\alpha \ast },X])
\label{eq:H}
\end{eqnarray}
where 
\begin{eqnarray}
\tilde{L}_{\alpha } &=&T_{\alpha k}(L_{k}+S_{kl}\beta _{l})+U_{\alpha
j}(R_{j}+\alpha _{j}) \\
\tilde{R}_{\alpha } &=&R_{j}\left[ U_{\beta k}N_{jk}+U_{\beta k}^{\ast
}M_{jk}^{\ast }\right] .
\end{eqnarray}
and with $K^{-1}=[K^{\alpha \beta }]$ the inverse matrix of $K=[K_{\alpha
\beta }]$, 
\begin{equation*}
\tilde{L}^{\alpha }=K^{\alpha \beta }\tilde{L}_{\beta },\quad \tilde{R}%
^{\alpha }=K^{\alpha \beta }\tilde{R}_{\beta }
\end{equation*}
and finally, the innovations processes are 
\begin{equation}
dI_{\alpha }(t)=dY_{\alpha }(t)-\pi _{t}(\tilde{L}_{\alpha }+\tilde{L}%
_{\alpha }^{\ast })\,dt.  \label{eq:innovations}
\end{equation}
\end{theorem}

Note that $\mathbb{E}[dI_{\alpha }(t)]\equiv 0$.

\subsection{Derivation of the Filter}

To derive a differential equation for the filter, we shall apply the
characteristic function technique to derive the filter for this problem. The
technique is to assume that the filter satisfies and equation of the form 
\begin{equation}
d\pi _{t}\left( X\right) =\mathcal{F}_{t}\left( X\right) dt+\mathcal{H}%
_{t}^{\alpha }\left( X\right) dY_{\alpha }\left( t\right)
\label{eq:filter_ansatz}
\end{equation}
where we assume that the processes $\mathcal{F}_{t}\left( X\right) $ and $%
\mathcal{H}_{t}^{j}\left( X\right) $ are adapted and lie in $\mathfrak{Y}%
_{t]}$.

To establish the proposition, we assume the form (\ref{eq:filter_ansatz})
and apply a method based on introducing a process $C\left( t\right) $
satisfying the QSDE 
\begin{equation}
dC\left( t\right) =f_{\alpha }\left( t\right) C\left( t\right) dY_\alpha
\left( t\right) ,
\end{equation}
with initial condition $C\left( 0\right) =I$. Here we assume that the $%
f_{\alpha }$ are integrable, but otherwise arbitrary. These coefficients may
be deduced from the identity 
\begin{equation*}
d\mathbb{E}\left[ \left( j_{t}\left( X\right) -\pi _{t}\left( X\right)
\right) C\left( t\right) \right] =0
\end{equation*}
which is valid since $C\left( t\right) \in \mathfrak{Y}_{t]}$. We note that
the It\={o} product rule implies $I+II+III=0$ where

\begin{eqnarray*}
I &=&\mathbb{E}\left[ \left\{ dj_{t}\left( X\right) -d\pi _{t}\left(
X\right) \right\} C\left( t\right) \right] , \\
II &=&\mathbb{E}\left[ \left( j_{t}\left( X\right) -\pi _{t}\left( X\right)
\right) dC\left( t\right) \right] , \\
III &=&\mathbb{E}\left[ \left( dj_{t}\left( X\right) -d\pi _{t}\left(
X\right) \right) dC\left( t\right) \right] .
\end{eqnarray*}

Here we have 
\begin{equation}
I=\mathbb{E}[\{j_{t}(\mathcal{L}X)-\mathcal{F}_{t}\left( X\right) -\mathcal{H%
}_{t}^{\alpha }\left( X\right) j_{t}(\tilde{L}_{\alpha }+\tilde{L}_{\alpha
}^{\ast })\}C\left( t\right) ]dt  \label{eq:I}
\end{equation}
\begin{equation*}
II=\mathbb{E}\left[ \left( \left\{ \pi _{t}\left( X\right) -j_{t}(X)\right\}
\right) j_{t}\left( \tilde{L}_{\alpha }+\tilde{L}_{\alpha }^{\ast }\right)
C\left( t\right) \right] f_{\alpha }\left( t\right) \,dt,
\end{equation*}
\begin{eqnarray*}
III &=&\,\mathbb{E}[\{j_{t}(\mathcal{L}_{0k}X)dB_{k}+j_{t}(\mathcal{L}%
_{jk}X)d\Lambda _{jk}+j_{t}\left( \mathcal{R}_{j0}X\right) dA_{j}^{\ast
}+j_{t}\left( \mathcal{R}_{0j}X\right) dA_{j} \\
&&-\mathcal{H}_{t}^{\beta }(X)dY_{\beta }\}\,dY_{\alpha }\,C(t)]\,f_{\alpha
}(t)
\end{eqnarray*}
To simplify term $III$ we note that 
\begin{eqnarray*}
j_{t}(\mathcal{L}_{0k}X)dB_{k}dY_{\alpha } &=&j_{t}(\mathcal{L}%
_{0k}X)\,T_{\alpha l}^{\ast }j_{t}(S_{lk}^{\ast })\,dt=j_{t}\left( \left[
L_{k}^{\ast },X\right] \right) T_{\alpha k}^{\ast }dt, \\
j_{t}(\mathcal{L}_{jk}X)d\Lambda _{jk}dY_{\alpha } &=&j_{t}(\mathcal{L}%
_{jk}X)T_{\alpha l}^{\ast }j_{t}\left( S_{lk}^{\ast }\right) dB_{j}^{\ast
}=j_{t}\left( [S_{lj}^{\ast },X]\right) T_{\alpha l}^{\ast }dB_{j}^{\ast },
\\
j_{t}(\mathcal{R}_{j0}X)dA_{j}^{\ast }dY_{\alpha } &=&j_{t}(\mathcal{R}%
_{j0}X)\,\left[ U_{\alpha k}N_{jk}+U_{\alpha k}^{\ast }M_{jk}^{\ast }\right]
\,dt=j_{t}\left( \left[ X,\tilde{R}_{\alpha }\right] \right) dt, \\
j_{t}(\mathcal{R}_{0j}X)dA_{j}dY_{\alpha } &=&j_{t}(\mathcal{R}_{0j}X)\,%
\left[ U_{\alpha k}M_{jk}+U_{\alpha k}^{\ast }(\delta _{jk}+N_{kj})\right]
\,dt \\
&=&j_{t}\left( \left[ R_{j}^{\ast },X\right] \right) U_{\alpha j}^{\ast
}dt+j_{t}\left( \left[ \tilde{R}_{\alpha }^{\ast },X\right] \right) dt,
\end{eqnarray*}
and so 
\begin{eqnarray*}
III &=&\mathbb{E}[j_{t}\left( \left[ \tilde{L}_{\alpha }^{\ast },X\right]
\right) C\left( t\right) ]f_{\alpha }(t)dt+\mathbb{E}[j_{t}\left( \left[ X,%
\tilde{R}_{\alpha }\right] \right) C\left( t\right) ]\,f_{\alpha }(t)dt \\
&&+\mathbb{E}[j_{t}\left( \left[ \tilde{R}_{\alpha }^{\ast },X\right]
\right) C\left( t\right) ]\,f_{\alpha }(t)dt-K_{\alpha \beta }\mathbb{E}[%
\mathcal{H}_{t}^{\beta }(X)C\left( t\right) ]f_{\alpha}(t)dt.
\end{eqnarray*}

Now from the identity $I+II+III=0$ we may extract separately the
coefficients of $f_{\alpha }\left( t\right) dt$ since the $f_{\alpha }$were
arbitrary. In particular we have $I\equiv 0$ as this is the only term not
proportional to $f_{\alpha }\left( t\right) dt$, and this implies that 
\begin{equation*}
\mathcal{F}_{t}\left( X\right) \equiv \pi _{t}(\mathcal{L}X)-\mathcal{H}%
_{t}^{\alpha }\left( X\right) \pi _{t}(\tilde{L}_{\alpha }+\tilde{L}_{\alpha
}^{\ast })
\end{equation*}

Similarly $II+III\equiv 0$ so 
\begin{eqnarray*}
0 &=&\mathbb{E}\left[ \left\{\pi _{t}(X\tilde{L}_{\alpha
}+X\tilde{L}_{\alpha }^{\ast })-\pi _{t}\left( X\right) \pi _{t}\left( \tilde{L%
}_{\alpha }+\tilde{L}_{\alpha }^{\ast }\right) \right\} C\left( t\right) \right] f_{\alpha
}\left( t\right) \,dt \\
&&+\mathbb{E}[\pi _{t}\left( \left[ \tilde{L}_{\alpha }^{\ast },X\right]
\right) C\left( t\right) ]f_{\alpha }(t)dt+\mathbb{E}[\pi _{t}\left( \left[
X,\tilde{R}_{\alpha }\right] \right) C\left( t\right) ]\,f_{\alpha }(t)dt \\
&&+\mathbb{E}[\pi _{t}\left( \left[ \tilde{R}_{\alpha }^{\ast },X\right]
\right) C\left( t\right) ]\,f_{\alpha }(t)dt-K_{\alpha \beta }\mathbb{E}[%
\mathcal{H}_{t}^{\beta }(X)C\left( t\right) ]f_{\alpha}(t)dt.
\end{eqnarray*}
From the coefficients of $f_{\alpha }\left( t\right) $ we deduce 
\begin{eqnarray*}
K_{\alpha \beta }\mathcal{H}_{t}^{\beta }(X) &=&\pi _{t}(X%
\tilde{L}_{\alpha }+\tilde{L}_{\alpha }^{\ast }X)-\pi _{t}\left( X\right) \pi
_{t}\left( \tilde{L}_{\alpha }+\tilde{L}_{\alpha }^{\ast }\right)  \\
&&+\pi _{t}\left( \left[ X,\tilde{R}_{\alpha }\right] \right) +\pi
_{t}\left( \left[ \tilde{R}_{\alpha }^{\ast },X\right] \right) ,
\end{eqnarray*}
which is the desired form.

\subsection{The Example}

\label{subsec:The Example} We now return to the specific model introduced in
the introduction. In this case we have $m_{\text{%
Fock}}=1,m_{\text{sq}}=1$ and $n_{\text{obs}}=2$. The system $G_{\text{sys}%
}\sim (S,L,H)$ is driven by the Fock input, while the squeezed input is
otherwise unprocessed $\left( R\equiv 0\right) $. We therefore have 
\begin{eqnarray*}
\tilde{L}_{1} &=&\frac{1}{\sqrt{2}}\left[ L+S\beta \left( t\right) \right] +%
\frac{1}{\sqrt{2}}\alpha \left( t\right) , \\
\tilde{L}_{2} &=&\frac{1}{\sqrt{2}i}\left[ L+S\beta \left( t\right) \right] -%
\frac{1}{\sqrt{2}i}\alpha \left( t\right) .
\end{eqnarray*}
The covariance matrix $K$ is now given by (\ref{eq: K}) and we find 
\begin{eqnarray*}
\tilde{L}^{1} &=&\frac{1+n-m^{\ast }}{\sqrt{2}\Delta }\left[ L+S\beta \left(
t\right) \right] +\frac{1+n-m}{\sqrt{2}\Delta }\alpha \left( t\right) , \\
\tilde{L}^{2} &=&\frac{1+n+m^{\ast }}{\sqrt{2}\Delta i}\left[ L+S\beta
\left( t\right) \right] -\frac{1+n+m}{\sqrt{2}\Delta i}\alpha \left(
t\right) ,
\end{eqnarray*}
with $\Delta =\left( 1+n\right) ^{2}-\left| m\right| ^{2}$, and of course $%
\tilde{R}^{1}=\tilde{R}^{2}=0$.

\section{Belavkin-Kalman Filters}

In this section we solve the filter problem for the special case of a linear
dynamical model with Gaussian state. For simplicity we take the system to be
a single cavity mode $a$ and have one squeezed input and (at most) one Fock
input.

Here it is possible to obtain an exact form for the filtered estimate $\pi
_{t}\left( a\right) $ in terms of the innovations. In particular, we will
encounter the following conditional covariances 
\begin{eqnarray*}
\mathscr{V}\left( t\right) &\triangleq &\pi _{t}\left( a^{\ast }a\right)
-\pi _{t}\left( a^{\ast }\right) \pi _{t}\left( a\right) , \\
\mathscr{W}\left( t\right) &\triangleq &\pi _{t}\left( a^{2}\right) -\pi
_{t}\left( a\right) ^{2}.
\end{eqnarray*}
It is a feature of the linear dynamics that these covariances will be
deterministic functions of time $t$. The assumption of gaussianity ensures
that higher mode moments can be written as combinations of products of first
and second order moments, so for instance 
\begin{eqnarray*}
\pi _{t}\left( XYZ\right) &=&\pi _{t}\left( X\right) \pi _{t}\left(
YZ\right) +\pi _{t}\left( Y\right) \pi _{t}\left( XZ\right) +\pi _{t}\left(
Z\right) \pi _{t}\left( XY\right) \\
&&-2\pi _{t}\left( X\right) \pi _{t}\left( Y\right) \pi _{t}\left( Z\right)
\end{eqnarray*}
where $X,Y,Z$ may be $a$ or $a^{\ast }$. In particular, we will make use of
the identities 
\begin{eqnarray}
\pi _{t}\left( a^{\ast }a^{2}\right) &=& \pi _{t}\left( a^{\ast }\right) \pi
_{t}\left( a^{2}\right) +2\pi _{t}\left( a\right) \pi _{t}\left( a^{\ast
}a\right) -2\pi _{t}\left( a^{\ast }\right) \pi _{t}\left( a\right) ^{2}
\nonumber \\
&=& \pi_t (a^\ast ) \mathscr{W} (t) + 2 \pi_t (a) \mathscr{V} (t) +\pi_t (a^\ast ) \pi_t (a)^2,
\label{eq:astar asquared} \\
\pi _{t}\left( a^{3}\right) &=&3\pi _{t}\left( a\right) \pi _{t}\left(
a^{2}\right) -2 \pi _{t}\left( a\right) ^{3}
\nonumber \\
&=& 3 \pi_t (a) \mathscr{W} (t)   +\pi_t (a)^3 .\label{eq:a cubed}
\end{eqnarray}

\subsection{Cavity mode driven by Fock input whose output is mixed with
squeezed noise}

This is the set-up described in subsection \ref{subsec:The Example} with 
\begin{equation*}
S=e^{i\phi },\quad L=\sqrt{\kappa }a,\quad H=\omega a^{\ast }a.
\end{equation*}

\begin{figure}[htbp]
\centering
\includegraphics[width=0.750\textwidth]{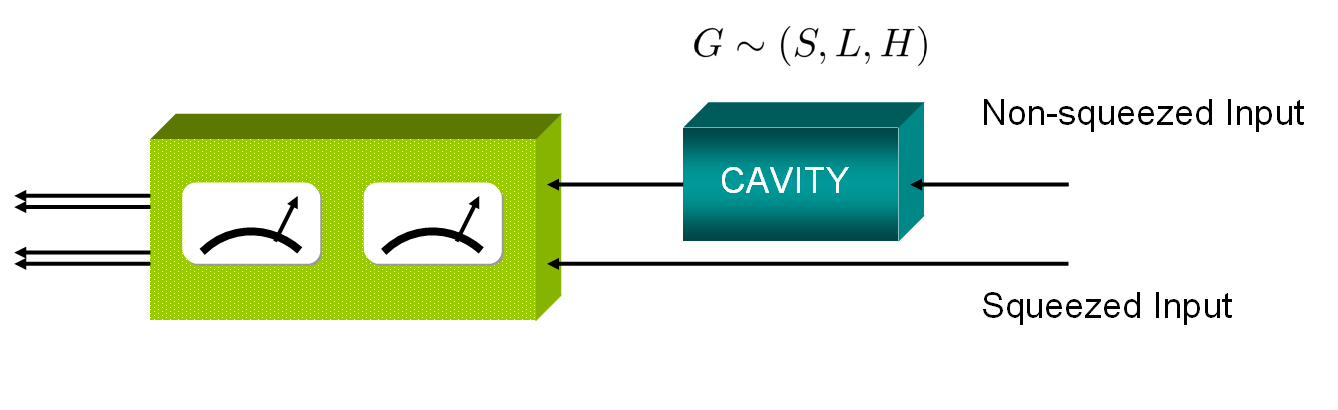}
\caption{System driven by a Fock input: the output subsequently mixed with a
squeezed input field.}
\label{fig:squeeze_mixed}
\end{figure}

(More generally we could take $L=\sqrt{\kappa _{-}}a+\sqrt{\kappa _{+}}%
a^{\ast }$ and $H=\omega a^{\ast }a+\varepsilon a^{\ast 2}+\varepsilon
^{\ast }a^{2}$ and still retain a linear dynamics: this is also solvable.)
We now have

\begin{eqnarray*}
\tilde{L}^{1} &=&\frac{1+n-m}{\sqrt{2}\Delta }\left[ \sqrt{\kappa }%
a+e^{i\phi }\beta \left( t\right) \right] +\frac{1+n-m^{\ast }}{\sqrt{2}%
\Delta }\alpha \left( t\right) , \\
\tilde{L}^{2} &=&\frac{1+n+m^{\ast }}{\sqrt{2}\Delta i}\left[ \sqrt{\kappa }%
a+e^{i\phi }\beta \left( t\right) \right] -\frac{1+n+m}{\sqrt{2}\Delta i}%
\alpha \left( t\right) .
\end{eqnarray*}
In the following section we will take $m$ to be real, that is, $m=m^{\prime }$
and $m^{\prime \prime }=0$. The filter equation then take the form 
\begin{eqnarray}
d\pi _{t}\left( X\right) &=&\pi _{t}(\frac{1}{2}\kappa a^{\ast }\left[ X,a%
\right] +\frac{1}{2}\kappa \left[ a^{\ast },X\right] a-i\omega \left[
X,a^{\ast }a\right]  \notag \\
&&+\sqrt{\kappa }\left[ X,a\right] e^{-i\phi }\beta ^{\ast }(t)+\sqrt{\kappa }%
\left[ a^{\ast },X\right] e^{i\phi }\beta (t))\,dt  \notag \\
&&+\sqrt{\frac{\kappa }{2\left( 1+n+m\right) }}\left\{ \pi _{t}\left(
Xa+a^{\ast }X\right) -\pi _{t}\left( X\right) \pi _{t}\left( a+a^{\ast
}\right) \right\} dW_{1}   \notag \\
&&+\frac{1}{i}\sqrt{\frac{\kappa }{2\left( 1+n-m\right) }}\left\{ \pi
_{t}\left( Xa-a^{\ast }X\right) -\pi _{t}\left( X\right) \pi _{t}\left(
a-a^{\ast }\right) \right\} dW_{2}  \notag \\
&&  \label{eq:filter_cav1}
\end{eqnarray}
where we now introduce rescaled processes 
\begin{equation*}
W_{1}\left( t\right) =\frac{1}{\sqrt{1+n+m}}I_{1}\left( t\right) ,\quad
W_{2}\left( t\right) =\frac{1}{\sqrt{1+n-m}}I_{2}(t).
\end{equation*}
The pair $W_{1}$ and $W_{2}$ are independent canonical Wiener processes and
we have 
\begin{equation*}
\left( dW_{1}\right) ^{2}=\left( dW_{2}\right) ^{2}=dt,\quad dW_{1}dW_{2}=0.
\end{equation*}

We obtain the following equation for the estimate $\pi _{t}\left( a\right) $ 
$:$%
\begin{eqnarray*}
d\pi _{t}\left( a\right) &=&-\left[ \left( \frac{1}{2}\kappa +i\omega
\right) \pi _{t}\left( a\right) +\sqrt{\kappa }e^{i\phi }\beta \left(
t\right) \right] dt \\
&&+\sqrt{\frac{\kappa }{2\left( 1+n+m\right) }}\left( \mathscr{W}\left(
t\right) +\mathscr{V}\left( t\right) \right) dW_{1}\left( t\right) \\
&&+\frac{1}{i}\sqrt{\frac{\kappa }{2\left( 1+n-m\right) }}\left( \mathscr{W}%
\left( t\right) -\mathscr{V}\left( t\right) \right) dW_{2}\left( t\right)
\end{eqnarray*}
We now compute $\mathscr{V}\left( t\right) $. We have
\begin{equation*}
d\pi _{t}\left( a^{\ast }a\right) =-\left[ \kappa \pi _{t}\left( a^{\ast
}a\right) +\sqrt{\kappa }\pi _{t}\left( a\right) e^{-i\phi }\beta \left( t\right)
^{\ast }+\sqrt{\kappa }\pi _{t}\left( a^{\ast }\right) e^{i\phi }\beta \left( t\right) %
\right] dt+\cdots .
\end{equation*}
Here, and in the following, expressions of the form $dP_{t}=G_{t}dt+\cdots $
retain drift terms, that is, the terms proportional to the increments of the
innovations are indicated by the ellipsis. From the It\={o} product rule,
\begin{equation*}
d\mathscr{V}\left( t\right) =d\pi _{t}\left( a^{\ast }a\right) -\left[ d\pi
_{t}\left( a^{\ast }\right) \,\pi _{t}\left( a\right) +\pi _{t}\left(
a^{\ast }\right) d\pi _{t}\left( a\right) +d\pi _{t}\left( a^{\ast }\right)
\,d\pi _{t}\left( a\right) \right]
\end{equation*}
Note that 
\begin{equation*}
d\pi _{t}\left( a^{\ast }\right) \,d\pi _{t}\left( a\right) =\Gamma \left(
t\right) \,dt
\end{equation*}
where 
\begin{eqnarray}
\Gamma \left( t\right) &=&\frac{\kappa }{2\left( 1+n+m\right) }\left| %
\mathscr{W}\left( t\right) +\mathscr{V}\left( t\right) \right| ^{2}+\frac{%
\kappa }{2\left( 1+n-m\right) }\left| \mathscr{W}\left( t\right) -\mathscr{V}%
\left( t\right) \right| ^{2}  \notag \\
&=&\frac{\kappa \left( 1+n\right) }{\Delta}\left( \mathscr{V}\left(
t\right) ^{2}+\left| \mathscr{W}\left( t\right) \right| ^{2}\right) -\frac{%
2\kappa m}{\Delta }\mathscr{V}\left( t\right) \,\text{Re\thinspace }%
\mathscr{W}\left( t\right)  \label{eq:Gamma_0}
\end{eqnarray}
and this leads to 
\begin{equation*}
d\mathscr{V}\left( t\right) =-\kappa \mathscr{V}\left( t\right) dt-\Gamma
\left( t\right) dt+\cdots .
\end{equation*}
The non-drift term in fact vanishes. To see this we would have to look at
the additional terms in $d\pi _{t}\left( a^{\ast }a\right) $ proportional to
the increments of the innovations. To this end we need to calculate $%
\mathcal{H}_{t}^{\alpha }\left( a^{\ast }a\right) $ which in this case
involves cubic terms $\pi _{t}\left( a^{\ast }a^{2}\right) ,\pi _{t}\left(
a^{\ast 2}a\right) ,\pi _{t}\left( a^{3}\right) $ and $\pi _{t}\left(
a^{\ast 3}\right) $. However, we may reduce this to first and second order
conditional expectations using (\ref{eq:astar asquared}) and (\ref{eq:a
cubed}). By inspection, the coefficients of $dW_{1}$ and $dW_{2}$ vanish
identically.

Similarly we have 
\begin{equation*}
d\mathscr{W}\left( t\right) =d\pi _{t}\left( a^{2}\right) -\left[ 2\pi
_{t}\left( a\right) d\pi _{t}\left( a\right) +d\pi _{t}\left( a\right)
\,d\pi _{t}\left( a\right) \right]
\end{equation*}
This time we have 
\begin{equation*}
d\pi _{t}\left( a\right) \,d\pi _{t}\left( a\right) =\Sigma \,\left(
t\right) dt
\end{equation*}
where 
\begin{eqnarray}
\Sigma \left( t\right) &=&\frac{\kappa }{2\left( 1+n+m\right) }\left( %
\mathscr{W}\left( t\right) +\mathscr{V}\left( t\right) \right) ^{2}-\frac{%
\kappa }{2\left( 1+n-m\right) }\left( \mathscr{W}\left( t\right) -\mathscr{V}%
\left( t\right) \right) ^{2}  \notag \\
&=&\frac{2\kappa \left( 1+n\right) }{\Delta}\mathscr{W}\left( t\right) \,\text{\thinspace }\mathscr{V}\left(
t\right) -\frac{\kappa m}{%
\Delta }\left( \mathscr{W}\left(
t\right) ^{2}+\mathscr{V}\left( t\right) ^{2}\right) .  \notag
\end{eqnarray}
This leads to the SDE 
\begin{equation*}
d\mathscr{W}\left( t\right) =-2\left(\frac{\kappa }{2}+i\omega \right)\mathscr{W}\left(
t\right) dt-\Sigma\left( t\right) \,dt+\cdots .
\end{equation*}
Once again, the cubic terms appearing may be replaced using (\ref{eq:astar
asquared}) and (\ref{eq:a cubed}) with the overall result that the drift
terms vanish identically. We can now give the explicit form for the filter.

\begin{proposition}
(With real squeezing parameter $m$.) The filtered estimate $\pi _{t}\left(
a\right) $ for the cavity mode satisfies 
\begin{eqnarray}
d\pi _{t}\left( a\right)  &=&-\left[ \left( \frac{1}{2}\kappa +i\omega
\right) \pi _{t}\left( a\right) +\sqrt{\kappa }e^{i\phi }\beta \left(
t\right) \right] dt  \notag \\
&&+\sqrt{\frac{\kappa }{2}}\frac{1}{\left( 1+n+m\right) }\left( \mathscr{W}%
\left( t\right) +\mathscr{V}\left( t\right) \right) dI_{1}\left( t\right)  
\notag \\
&&+\frac{1}{i}\sqrt{\frac{\kappa }{2}}\frac{1}{\left( 1+n-m\right) }\left( %
\mathscr{W}\left( t\right) -\mathscr{V}\left( t\right) \right) dI_{2}\left(
t\right)   \label{eq:cavity_a_filter}
\end{eqnarray}
where $\mathscr{V}\left( t\right) $ and $\mathscr{W}\left( t\right) $ are
deterministic functions satisfying the ODEs 
\begin{eqnarray}
\frac{d}{dt}\mathscr{V}\left( t\right)  &=&-\kappa \mathscr{V}\left(
t\right) -\frac{\kappa \left( 1+n\right) }{\Delta}\left( \mathscr{V}%
\left( t\right) ^{2}+\left| \mathscr{W}\left( t\right) \right| ^{2}\right)  
\notag \\
&&+\frac{2\kappa m}{\Delta }\mathscr{V}\left( t\right) \,\text{Re\thinspace }%
\mathscr{W}\left( t\right) ,  \label{eq:dV} \\
\frac{d}{dt}\mathscr{W}\left( t\right)  &=&-2\left(\frac{\kappa }{2}+i\omega \right)%
\mathscr{W}\left( t\right) -2\frac{\kappa \left( 1+n\right) }{\Delta}\mathscr{W}\left( t\right) \,\text{\thinspace }%
\mathscr{V}\left( t\right) 
   \notag \\
&&+\frac{\kappa m}{\Delta }\left( \mathscr{V}\left( t\right) ^{2}+\mathscr{W}\left( t\right)
^{2}\right) ,  \label{eq:dW}
\end{eqnarray}
with initial conditions $\mathscr{V}\left( 0\right) =\mathbb{E}\left[
a^{\ast }a\right] -\mathbb{E}\left[ a^{\ast }\right] \mathbb{E}\left[ a%
\right] $ and $\mathscr{W}\left( 0\right) =\mathbb{E}\left[ a^{2}\right] -%
\mathbb{E}\left[ a\right] ^{2}$.
The innovations are given by
\begin{eqnarray*}
dI_1 (t) &=&  dY_1 (t) -\sqrt{2}  \left[ 
\frac{ \sqrt{\kappa}}{2} \pi_t ( a+a^\ast ) +  \mathrm{Re} \{ e^{i\phi} \beta (t) 
+   \alpha (t) \} \right] dt \nonumber \\
dI_2 (t) &=&  dY_2 (t) -\sqrt{2} \left[ 
\frac{ \sqrt{\kappa}}{2} \pi_t \left( \frac{a-a^\ast}{i}\right) +  \mathrm{Im} \{ e^{i\phi} \beta (t) 
+  \, \alpha (t) \} \right] dt .
\end{eqnarray*}
\end{proposition}

The generalization to complex $m$ is straightforward, but algebraically more
involved due to the fact that the innovations are now correlated.

\subsection{Quantum filtering with direct squeezed input}

Our results also apply to the situation where we apply a squeezed field as
input to a system. In this can there can be no scattering. This falls into
general situation covered in Theorem 1 where we now ignore the Fock field $B$
(unless we wish to include further unmodelled dissipation). The output field
will of course be squeezed and we measure the quadrature 
\begin{equation*}
Y\left( t\right) =e^{-i\theta }A^{\text{out}}(t)+e^{i\theta }A^{\text{out}%
}\left( t\right)^{\ast} .
\end{equation*}
We therefore set $S=I,L=0$ with $R,$ the coupling operator to the single
(squeezed) input, nonzero.

\begin{figure}[htbp]
\centering
\includegraphics[width=0.750\textwidth]{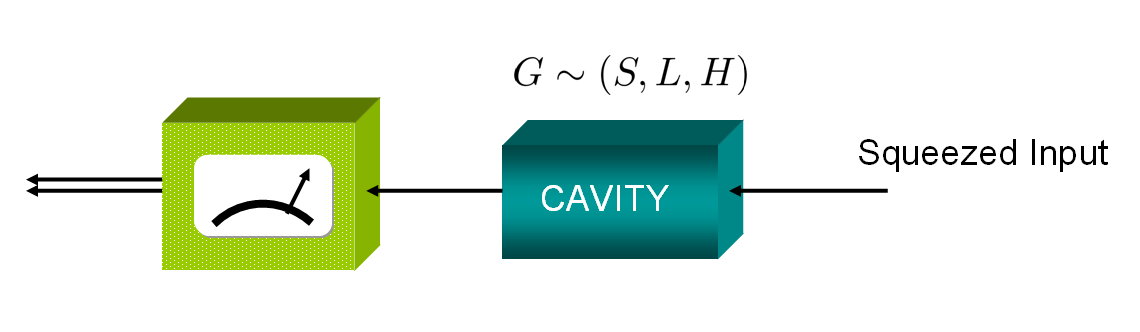}
\caption{System driven by a squeezed input (no scattering allowed!).}
\label{fig:squeeze_direct}
\end{figure}

The filter equation then simplifies to 
\begin{equation}
d\pi _{t}(X)=\pi _{t}(\mathcal{L}X)\,dt+\mathcal{H}_{t}\left( X\right)
dI\left( t\right) ,
\end{equation}
where 
\begin{equation*}
\mathcal{L}X=\mathcal{R}_{00}X+\alpha^{\ast }(t)\mathcal{R}_{10}X+%
\mathcal{R}_{01}X\alpha(t)-i\left[ X,H\right]
\end{equation*}
\begin{equation}
\mathcal{H}_{t}\left( X\right) =\pi _{t}(X\tilde{L}_{1}+\tilde{L}^{\ast }_{1}X)-\pi
_{t}(X)(\tilde{L}_{1}+\tilde{L}^{\ast }_{1})+\pi _{t}([X,\tilde{R}_{1}])+\pi _{t}([%
\tilde{R}_{1}^{\ast },X])
\end{equation}
where 
\begin{equation}
\tilde{L}_{1}=e^{i\theta }(R+\alpha \left( t\right) ),\quad \tilde{R}_{1}%
=\left( e^{i\theta }n+e^{-i\theta }m^{\ast }\right) R.
\end{equation}
and we require 
\begin{equation*}
\tilde{L}\equiv \frac{1}{K}\tilde{L}_{1},\quad \tilde{R}\equiv \frac{1}{K}%
\tilde{R}_{1}
\end{equation*}
with $K=1+2n+2\left( m^{\prime }\cos 2\theta +m^{\prime \prime }\sin 2\theta
\right)$. The innovations process is now 
\begin{equation}
dI(t)=dY(t)-\pi _{t}(\tilde{L}+\tilde{L}^{\ast })\,dt,
\end{equation}
and we $\left( dI\right) ^{2}=K\,dt$.

Again we specialize to the case of a cavity and take 
\begin{equation*}
R=\sqrt{\gamma }a,\quad H=\omega a^{\ast }a.
\end{equation*}

The filter equation is 
\begin{eqnarray*}
d\pi _{t}\left( X\right) &=&\frac{\gamma }{2}\left( 1+n\right) \pi
_{t}\left( \left[ a^{\ast },X\right] a+a^{\ast }\left[ X,a\right] \right)
\,dt \\
&&+\frac{\gamma }{2}n\pi _{t}\left( \left[ a,X\right] a^{\ast }+a\left[
X,a^{\ast }\right] \right) \,dt \\
&&-\frac{\gamma }{2}m^{\ast }\pi _{t}\left( \left[ a,X\right] a+a\left[ X,a%
\right] \right) \,dt \\
&&-\frac{\gamma }{2}m\pi _{t}\left( \left[ a^{\ast },X\right] a^{\ast
}+a^{\ast }\left[ X,a^{\ast }\right] \right) \,dt \\
&&-i\omega \pi _{t}\left( \left[ X,a^{\ast }a\right] \right) \,dt \\
&&+\frac{\sqrt{\gamma }}{K}\big\{ \pi _{t}(e^{i\theta }Xa+e^{-i\theta
}a^{\ast }X)-\pi _{t}(X)\pi _{t}(e^{i\theta }a+e^{-i\theta }a^{\ast }) \\
&& +\left( e^{i\theta }n+e^{-i\theta }m^{\ast }\right) \pi
_{t}([X,a])+\left( e^{-i\theta }n+e^{i\theta }m\right) \pi _{t}([a^{\ast
},X]) \big\} dI\left( t\right) .
\end{eqnarray*}

For $X=a$ we obtain 
\begin{eqnarray*}
d\pi _{t}\left( a\right)  &=&-\left( \frac{1}{2}\gamma +i\omega \right) \pi
_{t}\left( a\right) dt \\
&&+\frac{\sqrt{\gamma }}{K}\left\{ e^{-i\theta }\left( \mathscr{V}\left(
t\right) -n\right) +e^{i\theta }\left( \mathscr{W}\left( t\right) -m\right)
\right\} dI\left( t\right) .
\end{eqnarray*}
Observing that 
\begin{equation*}
d\pi _{t}\left( a^{\ast }\right) \,d\pi _{t}\left( a\right) =\frac{\gamma }{K%
}\left| e^{-i\theta }\left( \mathscr{V}\left( t\right) -n\right) +e^{i\theta
}\left( \mathscr{W}\left( t\right) -m\right) \right| ^{2}dt,
\end{equation*}
and 
\begin{eqnarray*}
d\pi _{t}\left( a^{\ast }a\right)  &=&\gamma (n-\pi _{t}\left( a^{\ast
}a\right) )\,dt \\
&&+\frac{\sqrt{\gamma }}{K}\{e^{i\theta }\pi _{t}\left( a^{\ast
}a^{2}\right) -e^{i\theta }\pi _{t}\left( a^{\ast }a\right) \pi _{t}\left(
a\right)  \\
&&+e^{-i\theta }\pi _{t}\left( a^{\ast 2}a\right) -e^{i\theta }\pi
_{t}\left( a^{\ast }a\right) \pi _{t}\left( a^{\ast }\right)  \\
&&-\left( e^{i\theta }n+e^{-i\theta }m^{\ast }\right) \pi _{t}\left(
a\right) -\left( e^{-i\theta }n+e^{i\theta }m\right) \pi _{t}(a^{\ast
})\}dI\left( t\right) 
\end{eqnarray*}
we see that 
\begin{eqnarray*}
d\mathscr{V}\left( t\right)  &=&d\pi _{t}\left( a^{\ast }a\right) -[d\pi
_{t}\left( a^{\ast }\right) \,\pi _{t}\left( a\right) +\pi _{t}\left(
a^{\ast }\right) d\pi _{t}(a)+d\pi _{t}\left( a^{\ast }\right) d\pi
_{t}\left( a\right) ] \\
&=&-\gamma \left( \mathscr{V}\left( t\right) -n\right) dt-\frac{\gamma }{K}%
\left| e^{-i\theta }\left( \mathscr{V}\left( t\right) -n\right) +e^{i\theta
}\left( \mathscr{W}\left( t\right) -m\right) \right| ^{2}dt,
\end{eqnarray*}
where we use the identity (\ref{eq:astar asquared}) again.

Similarly, we find 
\begin{equation*}
d\pi _{t}\left( a^{2}\right) =-2\left( \frac{1}{2}\gamma +i\omega \right)
\pi _{t}\left( a\right) \,dt+\gamma mdt+\cdots ,
\end{equation*}
and 
\begin{equation*}
\left( d\pi _{t}\left( a\right) \right) ^{2}=-\frac{\gamma }{K}\left\{
e^{-i\theta }\left( \mathscr{V}\left( t\right) -n\right) +e^{i\theta }\left( %
\mathscr{W}\left( t\right) -m\right) \right\} ^{2}dt,
\end{equation*}
so that 
\begin{eqnarray*}
d\mathscr{W}\left( t\right) &=&d\pi _{t}\left( a^{2}\right) -[2\pi
_{t}\left( a\right) d\pi _{t}(a)+d\pi _{t}\left( a\right) d\pi _{t}\left(
a\right) ] \\
&=&-2\left( \frac{1}{2}\gamma +i\omega \right) \mathscr{W}\left( t\right)
dt+\gamma mdt \\
&&-\frac{\gamma }{K}\left\{ e^{-i\theta }\left( \mathscr{V}\left( t\right)
-n\right) +e^{i\theta }\left( \mathscr{W}\left( t\right) -m\right) \right\}
^{2}dt.
\end{eqnarray*}

\begin{proposition}
The filtered estimate $\pi _{t}\left( a\right) $ for the cavity mode
satisfies 
\begin{eqnarray}
d\pi _{t}\left( a\right) &=&-\left( \frac{1}{2}\gamma +i\omega \right) \pi
_{t}\left( a\right) \,dt \notag \\
&&+\frac{\sqrt{\gamma }}{K}\left( e^{-i\theta }(\mathscr{V}\left( t\right)
-n)+e^{i\theta }(\mathscr{W}\left( t\right) -m)\right) d I(t)
\end{eqnarray}
where $\mathscr{V}\left( t\right) $ and $\mathscr{W}\left( t\right) $ are
deterministic functions satisfying the ODEs 
\begin{eqnarray}
\frac{d}{dt}\mathscr{V}\left( t\right) &=&-\gamma (\mathscr{V}\left(
t\right) -n)-\frac{\gamma }{K}\left| e^{-i\theta }(\mathscr{V}\left(
t\right) -n)+e^{i\theta }(\mathscr{W}\left( t\right) -m)\right| ^{2}, \\
\frac{d}{dt}\mathscr{W}\left( t\right) &=&-2i\omega \mathscr{W}-\gamma (%
\mathscr{W}\left( t\right) -m)  \notag \\
&&-\frac{\kappa }{2}\left( e^{-i\theta }(\mathscr{V}\left( t\right)
-n)+e^{i\theta }(\mathscr{W}\left( t\right) -m)\right) ^{2},
\end{eqnarray}
with initial conditions $\mathscr{V}\left( 0\right) =\mathbb{E}\left[
a^{\ast }a\right] -\mathbb{E}\left[ a^{\ast }\right] \mathbb{E}\left[ a%
\right] $ and $\mathscr{W}\left( 0\right) =\mathbb{E}\left[ a^{2}\right] -%
\mathbb{E}\left[ a\right] ^{2}$.
\end{proposition}

\bigskip \textbf{Acknowledgement} 
The authors would like to thank the Isaac Newton Institute for Mathematical Sciences, 
Cambridge, for support and hospitality during the programme \textit{Quantum Control Engineering} 
where work on this paper was completed.
JG is grateful
to the organizers of the meeting "Mathematical Aspects of Quantum Modeling,
Estimation and Control" in Padua, June 2013, where this work was begun, and
wishes to thank Sebastian Hofer for raising the squeezed filtering problem
there, and AD for support of London Mathematical Society scheme 3 grant.

\end{document}